\documentclass[english,aps,prl,twocolumn,amsmath,amssymb,showpacs,superscriptaddress,notitlepage,longbibliography]{revtex4-1}
\usepackage{bm}
\usepackage{hhline}
\usepackage{bbold}
\usepackage{makecell}
\usepackage{amsmath}
\usepackage{amssymb}
\usepackage{mathdots}
\usepackage{tabularx}
\usepackage{graphicx}
\usepackage{hyperref}
\usepackage{bbding}
\usepackage{tikz}
\usepackage{bbding} 
\usepackage{braket}
\usepackage{bm}

\makeatother

\usepackage{babel}
\begin{document}
\renewcommand{\figurename}{Fig.}
\title{Layer Pseudospin Superconductivity in Twisted MoTe$_2$}
	\author{Jin-Xin Hu}\thanks{Contact author: jhuphy@ust.hk}
        \affiliation{Department of Physics, Hong Kong University of Science and Technology, Clear Water Bay, Hong Kong, China}
  \author{Akito Daido}
  
       \affiliation{Department of Physics, Graduate School of Science, Kyoto University, Kyoto 606-8502, Japan}
       \affiliation{Department of Physics, Hong Kong University of Science and Technology, Clear Water Bay, Hong Kong, China}
   \author{Zi-Ting Sun}
	\affiliation{Department of Physics, Hong Kong University of Science and Technology, Clear Water Bay, Hong Kong, China}
   
    \author{Ying-Ming Xie}\thanks{Contact author: yxieai@connect.ust.hk}
    \affiliation{RIKEN Center for Emergent Matter Science (CEMS), Wako, Saitama 351-0198, Japan}
   \author{K. T. Law}\thanks{Contact author: phlaw@ust.hk}
   \affiliation{Department of Physics, Hong Kong University of Science and Technology, Clear Water Bay, Hong Kong, China} 	

\date{\today}
\begin{abstract}
Recent experiments have observed signatures of spin-valley-polarized unconventional superconductivity in twisted bilayer MoTe$_2$ (tMoTe$_2$). Here, we explore the rich physics of superconducting tMoTe$_2$, enabled by its unique layer-pseudospin structure. Within a minimal two-orbital layer-pseudospin model framework, both interlayer and intralayer Cooper pairings can be effectively visualized using a layer-space Bloch sphere representation. Remarkably, we find that interlayer pairing prevails in the spin-valley-polarized state, whereas intralayer pairing dominates in the spin-valley-unpolarized state. Strikingly, we further predict that for spin-valley-polarized intravalley superconducting state, experimentally feasible weak displacement fields can stabilize finite-momentum pairings at low temperatures. Additionally, in-plane magnetic fields, which break three-fold rotational symmetry, induce field-direction-dependent finite-momentum pairing states, leading to a versatile momentum-selection phase diagram. Our work highlights the crucial role of layer pseudospin in tMoTe$_2$'s unconventional superconductivity and demonstrates its unique tunability via external fields.
\end{abstract}
\maketitle


\section{Introduction}

Atomically thin layered quantum materials exhibit superconductivity following its initial observation in twisted bilayer graphene~\cite{cao2018unconventional,yankowitz2019tuning,lu2019superconductors,oh2021evidence}. Subsequently, other superconducting quasi-two-dimensional layered materials have been discovered, including Bernal bilayer graphene~\cite{zhou2022isospin,zhang2023enhanced,li2024tunable}, rhombohedral graphene multilayers~\cite{zhou2021superconductivity,choi2025superconductivity,han2025signatures} and twisted WSe$_2$~\cite{xia2025superconductivity,guo2025superconductivity}. These systems emerge as potential platforms for realizing topological superconductivity~\cite{xu2018topological,liu2018chiral,yu2021nematicity,xie2023gate,wu2019topological,gaggioli2025spontaneous,yang2024topological,qin2024chiral,chou2024intravalley,chatterjee2022inter,kim2025topological,christos2025finite}. Surprisingly, in a recent experiment rhombohedral tetralayer graphene supports strong evidence of chiral superconductivity~\cite{han2025signatures}, where the normal state is a spin-valley-polarized quarter-metal exhibiting anomalous Hall effect---a clear indication of broken time-reversal symmetry. The search for intrinsic chiral topological superconductors has attracted significant interests, as vortices in such systems can host Majorana zero modes~\cite{read2000paired,fu2008superconducting,law2009majorana}---exotic quasiparticles that obey non-Abelian statistics~\cite{kitaev2001unpaired,alicea2011non}.

Even more intriguingly, recently twisted bilayer MoTe$_2$ (tMoTe$_2$) have exhibited fractional Chern insulating states~\cite{cai2023signatures,park2023observation,zeng2023thermodynamic,xu2023observation,kang2024evidence} and possible signatures of superconductivity with clues of chiral nature~\cite{xu2025signatures,xu2025chiral}. This behavior arises because the superconducting state in tMoTe$_2$ emerges in adjacent to integer and fractional quantum anomalous Hall states, where magnetic switching behaviors and normal-state anomalous Hall effects are observed. These findings further support the potential for chiral superconductivity in tMoTe$_2$. The quantum anomalous Hall state (integer and fractional) in tMoTe$_2$ is expected to be spin-valley-polarized~\cite{wang2024fractional,reddy2023fractional,chen2025robust} due to the exceptionally strong Ising spin-orbit coupling (SOC)~\cite{xiao2012coupled,zhou2016ising,xie2022valley} and electron-electron interaction. Consequently, the superconducting pairing in such a system is more likely to be spin-triplet with odd-parity angular momentum to satisfy fermionic antisymmetry. Another key observation is that the superconductivity occurs within a narrow displacement field window ($\sim$1 meV), which is a characteristic energy scale comparable to the superconducting gap~\cite{xu2025signatures}. This suggests that the displacement field acts as a pair-breaking parameter, analogous to the role of a Zeeman field in disrupting Cooper pairs for conventional spin-singlet superconductors.

While previous theoretical studies have examined various aspects of superconductivity in twisted transition metal dichalcogenides (TMDs)~\cite{wu2023pair,qin2024kohn,xie2023orbital,zhu2025superconductivity,schrade2024nematic,kim2025theory,zhu2025plane}, the intriguing properties of spin-valley-polarized superconductivity observed in twisted MoTe$_2$ remain largely unexplored. In this work, motivated by the recent experimental discovery in tMoTe$_2$, we investigate the properties of spin-valley-polarized superconductivity from the perspective of the layer degree of freedom. 

\begin{figure}[t]
\centering
\includegraphics[width=1\columnwidth]{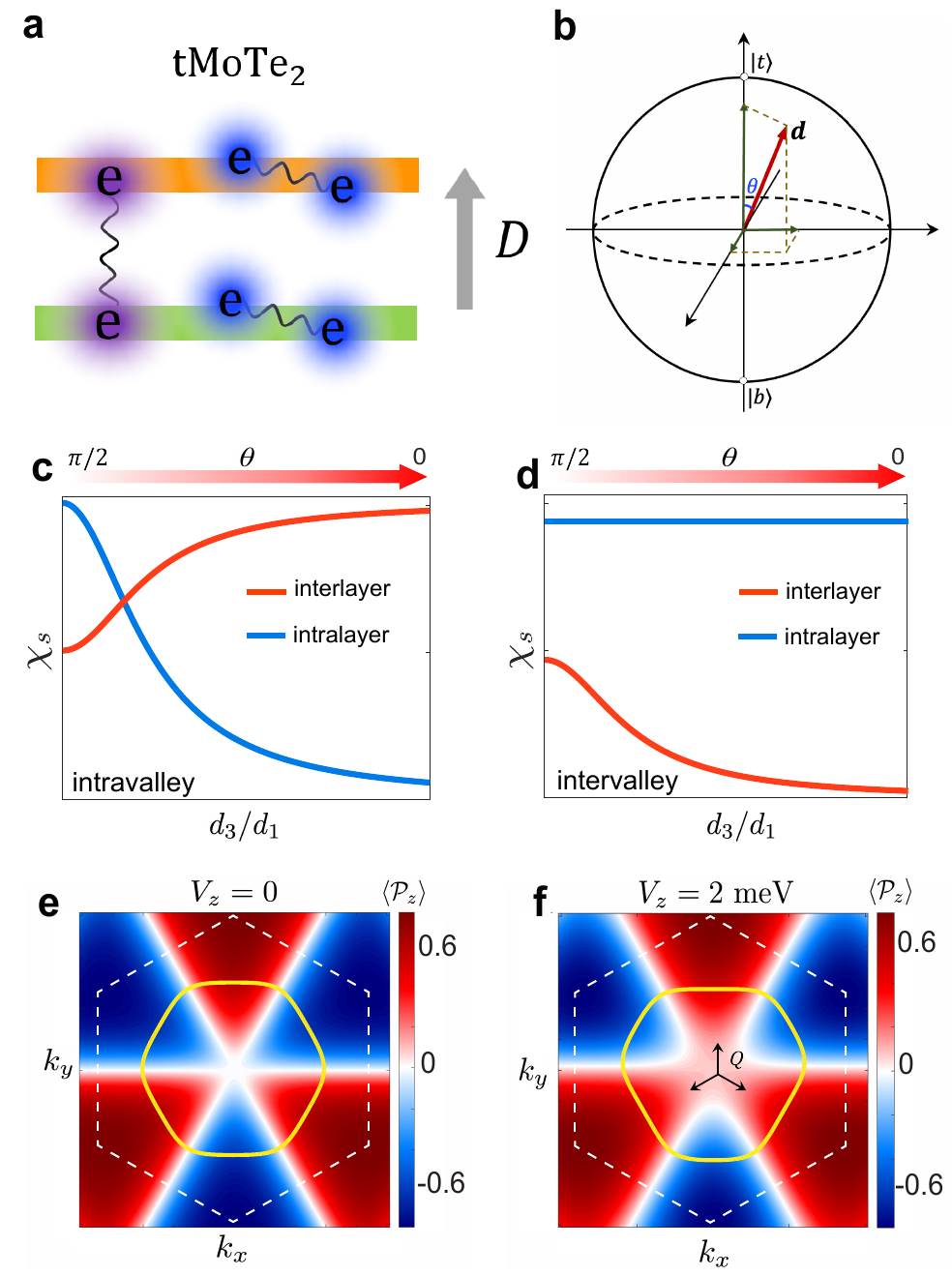}
\caption{\textbf{Interlayer and intralayer Cooper pairs in tMoTe$_2$.} \textbf{a} A schematic picture of the superconducting tMoTe$_2$ with both intralayer and interlayer Cooper pairs. \textbf{b} Bloch sphere in the layer-pseudospin basis. The minimal model Hamiltonian of twisted TMDs can be described by the four-component $d$ vector. \textbf{c},\textbf{d} The pairing susceptibility $\chi_s$. For $\theta\rightarrow 0$ (colatitude of $\bm{d}$ vector with respect to the $z$-axis), the interlayer pairing is prominent (minor) to the intralayer pairing for intravalley (intervalley) case. \textbf{e} The Fermi surface of 3.9$^\circ$-tMoTe$_2$ at hole filling $n_h=0.6$ (within $K$ valley) and the corresponding momentum-space layer polarization calculated from continuum model. \textbf{f} At finite displacement field, threefold-degenerate finite momentum pairing state can be stabilized.}
\label{fig:fig1}
\end{figure}

As illustrated in Fig.~\ref{fig:fig1}\textbf{a}, we consider that the bilayer system naturally hosts both interlayer and intralayer Cooper pairs. Despite the reduced Hilbert space due to Ising SOC, we reveal the role of the layer pseudospin and analyze the interlayer and intralayer pairings on a layer-space Bloch sphere (see Fig.~\ref{fig:fig1}\textbf{b}), with the model Hamiltonian described by the $\bm{d}$-vector: $H(\bm{k})=\bm{d}(\bm{k})\cdot \bm{\tau}$, where $\tau_j$ are Pauli matrices operating in layered space. Notably, the $\bm{d}$-vector's $z$-component ($d_3$) captures both the Fermi surface's hexagonal warping and opposite layer polarizations at opposite momenta within a single valley (see Fig.~\ref{fig:fig1}\textbf{e}). Because such unique layer polarization profile, the interlayer pairing would be strongly enhanced in the presence of intravalley pairing state, which happens in the spin-valley-polarized superconductors. With the layer polarization distributions within each valley, displacement fields would also cause the Fermi surface shifts, resulting in finite momentum pairings. Specifically, using this layer-pseudospin model, at weak displacement fields, we find that finite-momentum pairings become stable at low temperatures. We further demonstrate that the mixing of interlayer and intralayer Cooper pairs as well as finite momentum pairings from the layer-pseudospin model show qualitative agreement with the ones from the full continuum model. Finally, we propose that an in-plane magnetic field would break the three-fold degeneracy and lead to a phase transition from triple-$\bm{Q}$ to single-$\bm{Q}$ finite-momentum pairings.

Our work establishes a comprehensive framework for understanding the layer-pseudospin nature for the newly experimentally discovered spin-valley-polarized superconducting states in tMoTe$_2$ regardless of the specific pairing mechanism. We also propose a novel scheme for generating finite-momentum pairings through displacement fields, in sharp contrast to the conventional approach that relies on magnetic fields. 


\section{Results}

\textbf{Layer-pseudospin model.} To start with, we first introduce a minimal two-orbital layer-pseudospin model extended from the generalized Kane-Mele model~\cite{wu2019topological,devakul2021magic}. This model spans the two orbitals located at XM and MX stacking regions (see Methods for details), which effectively captures the low-energy physics of twisted TMDs and takes the following form
\begin{equation}
\label{eq:minimal_twoband}
H(\bm{k})=\bm{d}(\bm{k})\cdot\bm{\tau}+\frac{V_z}{2}\tau_z,
\end{equation}
where the four-component $\bm{d}$-vector [$\bm{d}=(d_0,d_1,d_2,d_3)$] are the hopping terms defined on a honeycomb lattice as $d_0=-2\gamma_0\sum_{j=1}^3 \cos(\bm{k}\cdot \bm{R}_j)$, $d_1-id_2=-\gamma_1\sum_{j=1}^3 e^{i\bm{k}\cdot \bm{a}_j}$ and $d_3=2\gamma_2\sum_{j=1}^3 \sin(\bm{k}\cdot \bm{R}_j)$. The lattice vectors are $\bm{a}_i=(\cos\frac{2(i-1)\pi}{3},\sin\frac{2(i-1)\pi}{3})L_M/\sqrt{3}$ and $\bm{R}_i=(\sin\frac{2(i-1)\pi}{3},\cos\frac{2(i-1)\pi}{3})L_M$ with $L_M$ the lattice constant. $\gamma_1$ describes the interlayer coupling and $\gamma_2$ contributes to the hexagonal warping of the Fermi surface and at small $\bm{k}$, $d_3=-\gamma_2 L_M^3 (k_y^3-3k_y k_x^2)/4+\mathcal{O}(k^4)$. This model successfully captures the topological character of the topmost moir\'{e} valence bands at small twist angles in twisted TMDs~\cite{wu2019topological,devakul2021magic}. A three-orbital tight binding model has recently been developed for improved accuracy at larger twist angles~\cite{qiu2023interaction,crepel2024bridging,kim2025theory}. In twisted TMDs, this model is for the single valley and the Hamiltonian of the other valley can be obtained as $H^*(-\bm{k})$. An out-of-plane displacement field $D$ (see Fig.~\ref{fig:fig1}\textbf{a}) breaks the $C_{2y}$ symmetry and introduces the layer potential difference $V_z$.

Interestingly, the momentum-space Hamiltonian of Eq.~\eqref{eq:minimal_twoband} in the Bloch basis $|\bm{k},t\rangle$ and $|\bm{k},b\rangle$ is characterized by the $\bm{d}$ vector on a layer Bloch sphere, as schematically depicted in Fig.~\ref{fig:fig1}\textbf{b}. The basis vector $|t\rangle=(1,0)^T$ and $|b\rangle=(0,1)^T$ denote the top and bottom layers, respectively. Therefore, the eigenstate of Eq.~\eqref{eq:minimal_twoband} are $| u_{\bm{k}}\rangle=(\cos\theta/2,e^{i\alpha}\sin\theta/2)^T$ (we focus on the upper band). Here $\alpha=\tan^{-1}(d_2/d_1)$ and 
\begin{equation}
\label{eq:layer_polarize}
\langle \mathcal{P}_z \rangle =\cos\theta=\frac{d_3}{|\bm{d}(\bm{k})|}.
\end{equation}
We also find that the layer polarization for the upper band $\langle \mathcal{P}_z \rangle \equiv \langle  u_{\bm{k}}|\tau_3| u_{\bm{k}}\rangle$ is directly controlled by $d_3$, which delineates the similar layer-polarization profile as $\langle \mathcal{P}_z\rangle  \sim \sin(3\theta_p)$ [the polar angle $\theta_p=\mathrm{tan}^{-1}(k_y/k_x)$] obtained from the continuum model (see Fig.~\ref{fig:fig1}\textbf{e}).

In the spin-valley-polarized superconducting state of tMoTe$_2$, we expect that $d_3$ would result in two essential features: (1) For intravalley pairing, even the total layer polarization is zero when the displacement field is not applied, the {\it layer-polarized} Fermi surface gives rise to the enhanced interlayer superconducting pairing. (2) At finite displacement fields, incommensurate finite-momentum pairing states could be favorable stemming from ``layer-Zeeman effect'', as shown in Fig.~\ref{fig:fig1}\textbf{d}. Note that in this case the zero momentum is defined as commensurate momentum $2\bm{K}$ because of the full-valley polarization~\cite{yang2024topological}.

\vspace{2mm}
\textbf{Intralayer and interlayer Cooper pairs.} To analyze the superconductivity in twisted TMDs, we employ the model interaction Hamiltonian on the first moir\'{e} band, which reads
\begin{equation}
H_{\mathrm{int}}=-\sum_{\bm{q}}U\Phi_0^\dagger(\bm{q})\Phi_0(\bm{q})+V\Phi_1^\dagger(\bm{q})\Phi_1(\bm{q}),
\end{equation}
where 
\begin{equation}
\Phi_i(\bm{q})=\sum_{\bm{k},\sigma\sigma'}\varphi_{\bm{k}}\Lambda_{i}^{\sigma\sigma'}(\bm{k},\bm{q}) c_{\bm{k}+\frac{\bm{q}}{2},\sigma}c_{-\bm{k}+\frac{\bm{q}}{2},\sigma'}.
\end{equation}
Here $\bm{q}$ is the Cooper pair momentum, $\varphi_{\bm{k}}$ is the angular structure of concrete pairing and $\sigma,\sigma'$ denote the spin/valley index. The form factor $\Lambda_{i}^{\sigma\sigma'}(\bm{k},\bm{q})=\langle u_{\bm{k}+\frac{\bm{q}}{2},\sigma}|\tau_i|u^*_{-\bm{k}+\frac{\bm{q}}{2},\sigma'}\rangle$ are the overlap functions for the different pairing channels. $U,V$ denote the strength of intralayer and interlayer interactions and $c_{k}$ is the band-basis electron annihilation operator. Motivated by the recent experiment~\cite{xu2025signatures}, we assume $\varphi_{\bm{k}}$ is the chiral $p$-wave pairing in the spin-valley-polarized state with $\sigma=\sigma'$. For illustration throughout this work we the pairing function $\varphi_{\bm{k}}=\sum_{j=1}^3 \omega^{j-1} \sin(\bm{k}\cdot \bm{R}_j)$ with $\omega=\mathrm{exp}(-2i\pi/3)$. For intervalley pairing, $\sigma=-\sigma'$ and $\varphi_{\bm{k}}=1$ as the $s$-wave pairing.


\begin{figure}
		\centering
		\includegraphics[width=1\linewidth]{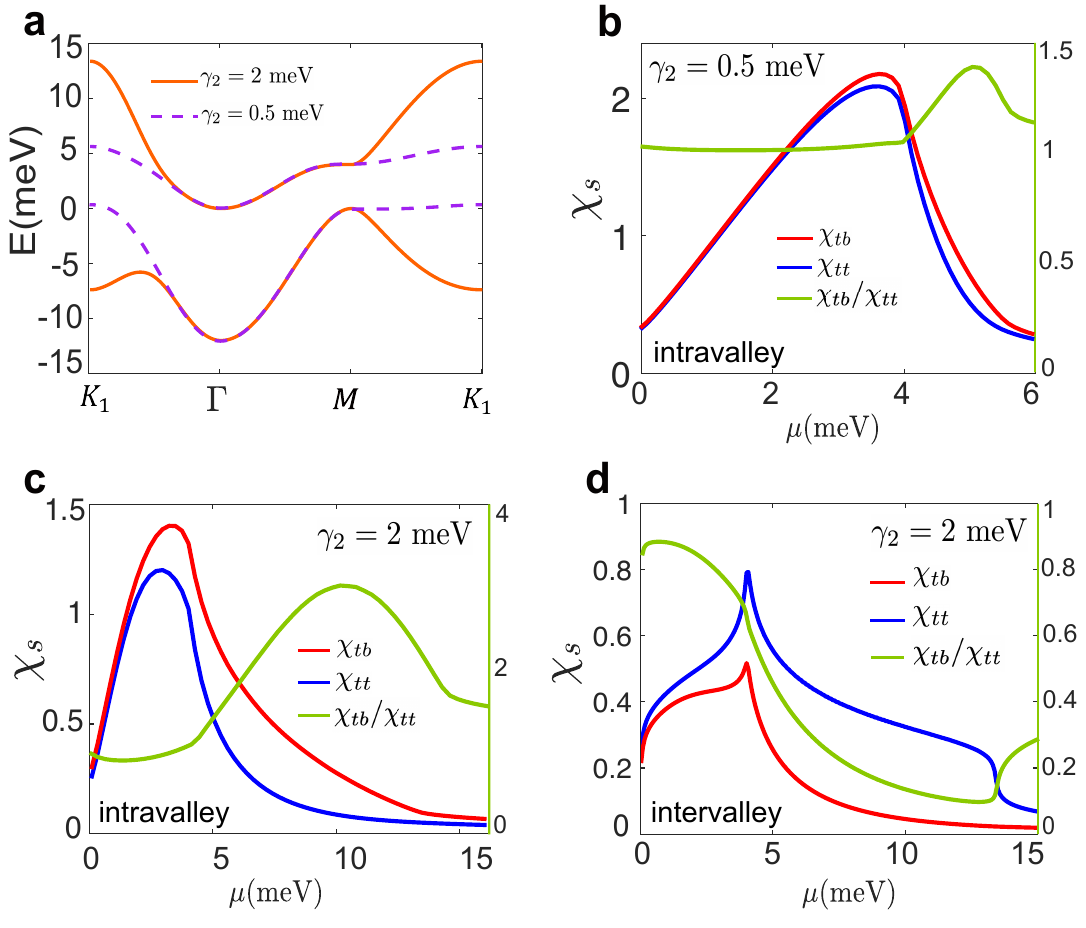}
		\caption{\textbf{Enhanced interlayer pairing for layer-pseudospin model.} \textbf{a} The band structure of layer-pseudospin model in Eq.~\eqref{eq:minimal_twoband} at $\gamma_2=2$meV (solid orange lines) and $\gamma_2=0.5$meV (dashed purple lines). Other parameters: $(\gamma_0,\gamma_1)=(1,2)$meV. \textbf{b}, \textbf{c} Superconducting pairing susceptibility in the intralayer ($\chi_{tt}$) and intralayer ($\chi_{tb}$) channels as a function of the Fermi energy $\mu$. \textbf{d} For intervalley pairing, $\chi_{tt}$ dominate over $\chi_{tb}$ as a comparison of \textbf{c}. For all calculations, the temperature is set to be $k_BT=0.02$meV.}
		\label{fig:fig2}
\end{figure}

To explore the pairing tendency towards different channels, we analyze the the pairing susceptibility 
\begin{equation}\label{sus}
\chi_{ij}(\boldsymbol{q})=\sum_{\boldsymbol{k},\sigma\sigma'} \mathcal{K}_{\sigma\sigma'}(\bm{k},\bm{q})| \varphi_{\bm{k}}|^2 \Lambda_{i}^{\sigma\sigma'}(\bm{k},\bm{q})\Lambda_{j}^{\sigma\sigma'*}(\bm{k},\bm{q}).
\end{equation}
with the kernel function
\begin{equation}
\mathcal{K}_{\sigma\sigma'}(\bm{k},\bm{q})=\frac{\tanh\frac{\xi_{\boldsymbol{k+q/2},\sigma}}{2k_B T}+\tanh\frac{\xi_{\boldsymbol{-k+q/2},\sigma'}}{2k_B T}}{2(\xi_{\boldsymbol{k+q/2},\sigma}+\xi_{\boldsymbol{-k+\bm{q}/2},\sigma'})}
\end{equation}
Here $\xi_{\bm{k},\sigma}=\varepsilon_{\bm{k},\sigma}-\mu$ with $\varepsilon_{\bm{k},\sigma}$ the bare band dispersion and $i$ the channel index. $k_B$ is the Boltzmann's constant, and $T$ is the temperature. In twisted bilayer TMDs, both interlayer and intralayer Cooper pairs are involved, depending on the microscopic interlayer and intralayer interactions ($V$ and $U$). Assuming comparable interaction strengths $U\approx V$, we determine the dominant pairing channel by comparing the corresponding susceptibilities. The superconducting transition temperature is determined by the full pairing susceptibility matrix (see methods). In the following contents, we use $\chi_{tt}$ and $\chi_{tb}$ to denote the intralayer ($\chi_{00}$) and interlayer ($\chi_{11}$) pairings, respectively. As we will see later, even though the two pairing channels mix in general cases, the interlayer one dominates over the intralayer one for the intravalley superconducting pairing. 


Fig.~\ref{fig:fig2}\textbf{a} displays the single-particle band structure obtained from Eq.~\eqref{eq:minimal_twoband} with parameters $(\gamma_0,\gamma_1)=(1,2)\mathrm{meV}$. The zero-momentum pairing susceptibility $\chi_s$ is presented in Fig.~\ref{fig:fig2}\textbf{b}, where $\chi_{tt}$ and $\chi_{tb}$ denote the intralayer and interlayer pairing susceptibilities respectively [$i=0,1$ in Eq.~\eqref{sus}]. For $\gamma_2=2$ meV,  $\chi_{tb}/\chi_{tt}$ shows significant enhancement at $\mu>5$ meV, coinciding with increased layer polarization associated with the Fermi surface in this regime. At smaller $\gamma_2=0.5$meV we find that $\chi_{tt}\approx \chi_{tb}$ across all $\mu$ values, indicating nearly equal mixing of intra- and interlayer pairings due to the weaker $d_3$ term. In sharp contrast, we also show the case for intervalley pairing with the pairing function $\varphi_{\bm{k}}=1$. We can always find that $\chi_{tt}> \chi_{tb}$ suggesting the intralayer pairing dominates in this case and $\chi_{tb}/\chi_{tt}$ decreases upon increasing $\mu$.

To understand the discrepancy of $\chi_{tt}$ and $\chi_{tb}$, we evaluate the zero-momentum pairing susceptibility for the layer-pseudospin model. Focussing on the intravalley pairing ($\sigma=+$), the overlap functions are given by $|\langle u_{\bm{k},+}|\tau_x|u^*_{-\bm{k},+}\rangle|^2=1-\sin^2\theta(1/2-\cos 2\alpha)$ and $|\langle u_{\bm{k},+}|\tau_0|u^*_{-\bm{k},+}\rangle|^2=\sin^2\theta$. Therefore, we can obtain 
\begin{equation}
\label{eq:sus_inter}
\left\{
\begin{aligned}
 &\chi_{tb}=\sum_{\bm{k}}[1-\sin^2\theta(1/2-\cos 2\alpha)]|\varphi_{\bm{k}}|^2\mathcal{K}_{++}(\bm{k},0) \\   
 &\chi_{tt}=\sum_{\bm{k}}\sin^2\theta|\varphi_{\bm{k}}|^2 \mathcal{K}_{++}(\bm{k},0),
\end{aligned}
\right.
\end{equation}
In the limit of $\gamma_2\gg \gamma_1$, $\theta \rightarrow 0$ leads to $\chi_{tb}\rightarrow \sum_{\bm{k}}|\varphi_{\bm{k}}|^2 \mathcal{K}_{++}(\bm{k},0)$ and $\chi_{tt}\rightarrow 0$, where the interlayer coupling is weakest. In this regime, the interlayer pairing is prominent and the intralayer pairing can be neglected. In contrast, in the limit of $\gamma_2\rightarrow 0$, $\theta \rightarrow \pi/2$ such that $\chi_{tb}\rightarrow \sum_{\bm{k}}|\varphi_{\bm{k}}|^2 \cos^2(\alpha/2) \mathcal{K}(\bm{k},0)$ and $\chi_{tt}\rightarrow \sum_{\bm{k}}|\varphi_{\bm{k}}|^2 \mathcal{K}(\bm{k},0)$. In this case, $\chi_{tt}$ and $\chi_{tb}$ become comparable in magnitude.

For intervalley pairing state, $|\langle u_{\bm{k},+}|\tau_x|u_{-\bm{k},-}\rangle|^2=\sin^2\theta\cos^2\alpha$ and $|\langle u_{\bm{k},+}|\tau_0|u_{-\bm{k},-}\rangle|^2=1$. Therefore, the intralayer channel is always stronger than interlayer channel, and in the limit of $\gamma_2\gg \gamma_1$, $\sin\theta\rightarrow 0$ renders $\chi_{tb}\rightarrow 0$. Based on the above pairing susceptibility analysis on a layer Bloch sphere, we schematically summarize how the interlayer and intralayer Cooper pairs get influenced by the layer pseudospin structure in Fig.~\ref{fig:fig1}\textbf{c} and \textbf{d}. Hence, we conclude that the mixing between interlayer and intralayer pairings are quite different for intervalley and intravalley superconductivity. 
\begin{figure}
		\centering
		\includegraphics[width=1\linewidth]{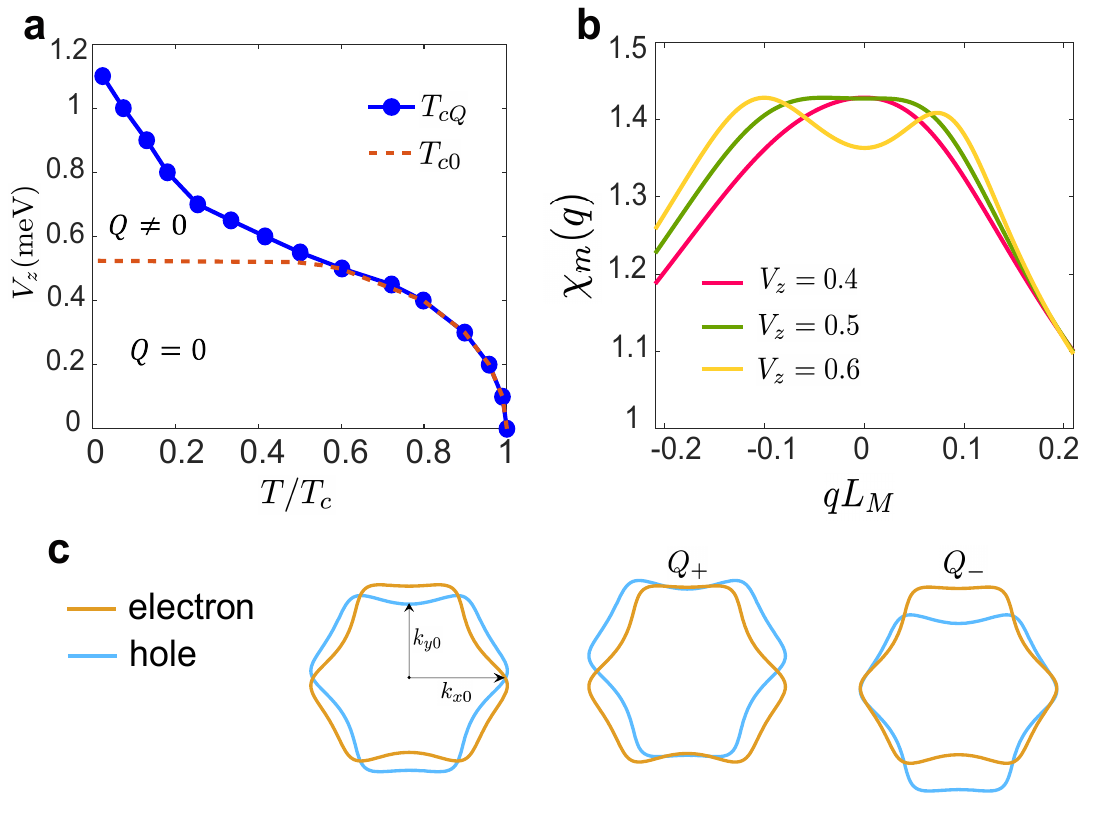}
		\caption{\textbf{Displacement field induced finite momentum pairings for layer-pseudospin model.} \textbf{a} The phase diagram of finite momentum pairing state. The finite momentum $Q\neq 0$ is stabalized at finite $V_z$ and low temperatures. \textbf{b} The largest eigenvalue of pairing susceptibility matrix $\chi_m(q)$ at different gate field $V_z$. In the calculations we set $U=V=0.7$ meV and $\mu=4$meV, which yields $k_B T_c\approx 0.1$meV. \textbf{c} Electron and hole Fermi surfaces with finite displacement field. There are two nondegenerate nesting vectors $Q_\pm$ for the finite momentum pairings. $k_{x0},k_{y0}$ are Fermi momenta along $x$ and $y$ directions. 
        }
		\label{fig:fig3}
\end{figure}

\vspace{2mm}
\textbf{FFLO phase through layer-Zeeman effect.} Having studied the interlayer and intralayer pairings for zero-momentum pairing states in the previous section, we now move to the possible finite-momentum pairing states under finite displacement fields. 

To be concrete, we solve the linearized gap equation to obtain the $V_z-T$ phase diagram. In Fig.~\ref{fig:fig3}\textbf{a} we plot the calculated critical temperature $T_{c\bm{Q}}$ and $T_{c0}$. We can observe that finite momentum pairing states $\bm{Q}\neq 0$ emerge at low temperatures and above a critical field $V_c$, exhibiting an upturn at the phase transition.  This phase diagram bears strong analogy to the FFLO phase in conventional superconductors, with the displacement field $V_z$ playing a role analogous to the magnetic field~\cite{fulde1964superconductivity,larkin1965nonuniform,liu2017unconventional,xie2023orbital,wan2023orbital}. We term this mechanism the {\it layer-Zeeman effect}. To further elucidate the phase transition, in Fig.~\ref{fig:fig3}\textbf{b} we depict the largest eigenvalue of pairing susceptibility matrix $\chi_m(q)$ near the critical temperature upon increasing the gate field $V_z$. In general there are two nondegenerate minima at $\bm{Q}_+$ and $\bm{Q}_-$. Because of the $C_{3z}$ symmetry, $\bm{Q}$ has three-fold degeneracy so we choose the one $\bm{Q}=(0,1)Q$ for illustration.


The phase diagram in Fig.~\ref{fig:fig3}\textbf{a} delineates the general picture of layer-Zeeman effect. 
To further understand the upper critical field $V_c$ and the field-induced FFLO phase, we model the band dispersion as $\varepsilon_k=\sqrt{v_0^2\bm{k}^2+(\lambda k^3\sin3\theta_p+V_z/2)^2}$~\cite{fu2009hexagonal}, where $v_0$ and $\lambda$ are free parameters. $\lambda$ encodes Fermi-surface hexagonal warping. Similar to the Zeeman effect, the upper critical field $V_c$ is given by $V_c\propto \sqrt{T_c-T}/\lambda$. The gate tunable finite momentum can be understood by the Fermi surface nesting as depicted in Fig.~\ref{fig:fig3}\textbf{c}. The two nesting vectors $Q_+$ and $Q_-$ correspond to the coincidence of the electron and hole Fermi surfaces. Analytically, we can obtain 
\begin{equation}
Q_+\approx\frac{V_z}{3\lambda k_{x0}^2}, Q_-\approx-\frac{V_z\lambda k_{y0}^2}{3\lambda^2 k_{y0}^4+2v_0^2},
\end{equation}
where $k_{x0}$, $k_{y0}$ are Fermi momenta at $x$ and $y$ directions (see Fig.~\ref{fig:fig3}\textbf{c}). Such a phenomenological analysis reveals how the displacement field and hexagonal warping affect the upper critical field and finite-momentum pairings.

\vspace{2mm}

\textbf{Results from continuum model.} To further strengthen the results obtained from our two-orbital layer pseudospin model, we now turn to the continuum-model description of twisted TMDs~\cite{wu2019topological}. To capture the essential physics, we adopt the parameters from Ref.~\cite{xu2024maximally}, despite the fact that several DFT studies of tMoTe$_2$ exist with alternative parameter sets~\cite{read2000paired,jia2024moire,wang2024fractional,zhang2024polarization}.
\begin{figure*}
		\centering
		\includegraphics[width=1\linewidth]{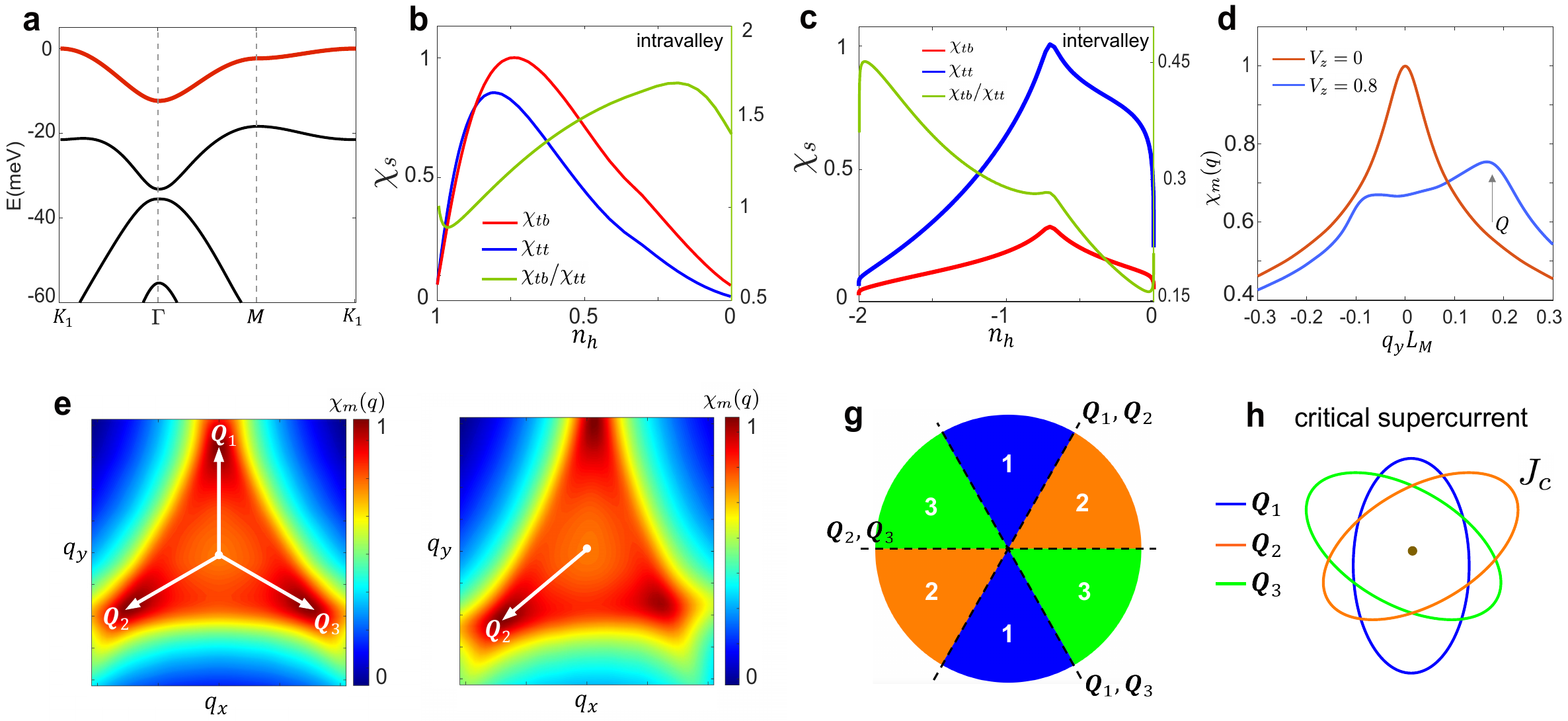}
		\caption{\textbf{Results for continuum model of tMoTe$_2$.} \textbf{a} The moir\'{e} band structure of tMoTe$_2$ at $\theta_t=3.9^\circ$ with zero displacement field. \textbf{b} Renormalized superconducting pairing susceptibility $\chi_s$ for interlayer ($\chi_{tb}$) and intralayer ($\chi_{tt}$) channels. For intravalley pairing, $\chi_{tb} > \chi_{tt}$ when $n_h<0.87$. For intervalley spin-singlet pairing (with $s$-wave basis function $\varphi_{\bm{k}}=1$), $\chi_{tt} \gg \chi_{tb}$. \textbf{d} For intravalley pairing, $\chi_m(q)$ has two peaks at finite displacement field. Finite momentum pairings are stablized at nondegenerate momentum $Q_{+}$. The temperature is set to be $k_BT=0.02$meV and $n_h=0.7$. \textbf{e} The landscape of $\chi_m(q)$ shows the minimum at three degenerate momentum $\bm{Q}_{1,2,3}$. They are connected by $C_{3z}$. \textbf{f} In the presence of in-plane magnetic field with the polar angle $\theta_B=\pi/6$. In this case, the $\bm{Q}_2$ state has the lowest energy. \textbf{g} The favored momentum goes as $\bm{Q}_2-\bm{Q}_1-\bm{Q}_3$ with respect to the polar angle. \textbf{h} The critical supercurrent $J_c$ has anisotropy for different finite momentum state.}
		\label{fig:fig4}
\end{figure*}

Figure~\ref{fig:fig4}\textbf{a} shows the moir\'{e} band structure for $K$-valley electrons of tMoTe$_2$ with the topmost valence band featuring Chern number $C=1$. As we focus on the superconductivity within single $K$ valley, from the pairing susceptibility calculation, in Fig.~\ref{fig:fig4}\textbf{b} we can find that $\chi_{tb}> \chi_{tt}$ as carrier density goes up, implying the interlayer pairing is stronger than the intralayer pairing for the intravalley case. This is consistent with Fig.~\ref{fig:fig2}\textbf{c}. Notice that the peak in $\chi_s$ is not at Van Hove singularity  because of the momentum-dependent pairing function $\varphi_{\bm{k}}$. On the other hand, for intervalley pairing $\chi_{tt}\gg \chi_{tb}$ as displayed in Fig.~\ref{fig:fig4}\textbf{c}, which is consistent with resent works~\cite{xie2023orbital,zhu2025superconductivity}. These observations can be understood from layer-pseudospin mechanism in Eq.~\eqref{eq:sus_inter} and Fig.~\ref{fig:fig1}\textbf{c} and \textbf{d}. We would like to emphasize that while the layer-pseudospin analysis offers valuable qualitative insights, it does not achieve quantitative agreement with our continuum model calculations.

The finite-momentum pairing states can be achieved by applying finite displacement field that introduces layer potential $V_z$. In the experiment~\cite{xu2025signatures}, the superconductivity is suppressed near $E_z\approx 15$ mV/nm, corresponding to $V_z\approx 1.2$ meV. To show the finite-momentum pairing instabilities arising from Fermi surface nesting, in Fig.~\ref{fig:fig4}\textbf{d} we plot $\chi_m(q)$. It can be seen that even for a weak displacement field $V_z=0.8$ meV (experimentally achievable), $Q$ remains sizable, reaching $\sim 0.2 L_M^{-1}\approx 0.04\mathrm{nm}^{-1}$. Therefore, in experiments, we expect that the finite-momentum pairing states could be manifested by a upturn in the $V_z-T$ phase diagram that would serve as a direct experimental signature. Additionally, due to the $C_{3z}$ symmetry, there are three degenerate states $\bm{Q}_1,\bm{Q}_2,\bm{Q}_3$ as shown in Fig.~\ref{fig:fig4}\textbf{e}.

\vspace{2mm}
\textbf{Finite-momentum selection by in-plane magnetic field.} Under displacement field, the layer pseudospin gives rise to the chiral triple-$\bm{Q}$ pairing state because of rotation symmetry and intravalley pairing nature.  This gate-tunable exotic pairing state naturally stabilizes into an energetically favorable vortex-antivortex lattice, which hosting Majorana zero modes~\cite{gaggioli2025spontaneous}. However, this three-fold degeneracy may be lifted by intrinsic or extrinsic perturbations, such as nematicity or uniaxial strain~\cite{hu2022nonlinear,daido2025current}. We here propose that an in-plane magnetic field can externally control the energetics of finite-momentum states via orbital-layer coupling, with the field orientation serving as a tuning knob. Specifically, for an in-plane magnetic field $\bm{B}=B(\cos\theta_B,\sin\theta_B,0)$, the vector potential $\bm{A}=d\bm{B}\times \hat{z}/2$ imposes a layer-dependent momentum shift $\delta\bm{k}=\pm Bd(\sin\theta_B,-\cos\theta_B)$  where the $+(-)$ sign corresponds to the top (bottom) layer. Here $\theta_B$ deonotes the polar angle of the magnetic field. 

As shown in Figs.~\ref{fig:fig4}\textbf{e}-\textbf{f}, applying an in-plane magnetic field $B$ can drive a first-order transition from a triple-$\bm{Q}$ state to a single-$\bm{Q}$ state.  This transition occurs through a redistribution of $\chi_{m}(q)$, lowering it at $\bm{Q}_1$ and $\bm{Q}_3$, while raising it at $\bm{Q}_2$. Here we assume that the magnetic field is not strong enough to break the superconductivity and the positions of $\bm{Q}$ are approximately unchanged. Note that, as discussed in previous sections, the finite momentum $Q$ arises from displacement fields. Perhaps more interestingly, rotating the B-field direction reveals a rich phase diagram as displayed in Fig.~\ref{fig:fig4}\textbf{f}. When $B$ aligns with the three dashed lines in Fig.~\ref{fig:fig4}\textbf{f} (i.e., at angles $\theta_B=n\pi/3$ with integer $n$), double-$\bm{Q}$ states emerge due to energy degeneracy between two momenta. For instance, at $\theta_B=\pi/3$, $\bm{Q}_1$ and $\bm{Q}_2$ become degenerate. To probe these first-order transitions experimentally, one feasible approach is to measure the critical supercurrent anisotropy. Figure~\ref{fig:fig4}\textbf{h} illustrates this anisotropy: the angle-dependent critical supercurrent $J_c$ varies distinctly for different momentum $\bm{Q}$ (purple ellipse), reflecting the underlying finite-momentum pairing states. It is then expected that the first-order transitions induced by the in-plane magnetic field should appear as discontinuous jumps in the critical supercurrents. Because of the low symmetry, superconducting diode effect is also expected in this regime.

\section{Discussion}
In van der Waals-stacked layered quantum materials, electronic properties are fundamentally coupled to the layer degree of freedom. The emerging concept of ``layertronics'' in twisted bilayer systems exploits the interplay between twist angle, interlayer coupling, and moir\'{e} physics to engineer new quantum states of matter and functional electronic devices~\cite{gao2020tunable,gao2021layer,zhai2023time,fan2024intrinsic,zhang2025experimental}. The observed superconductivity in twisted WSe$_2$~\cite{xia2025superconductivity,guo2025superconductivity} and MoTe$_2$~\cite{xu2025signatures} have expanded the scope of superconducting layertronics in moir\'{e} systems.

In our work, we report the key findings and implications for layer pseudospin superconductivity in tMoTe$_2$ through a two-orbital layer-pseudospin model. Unlike the previous works focusing on intervalley pairing, the dominant intravalley pairing shows its exclusive behaviors in tMoTe$_2$. Our minimal model unveils the fundamental role of layer-pseudospin in the mixing between interlayer and intralayer pairings and the emergence of finite-momentum pairing instabilities under displacement fields. Our work establishes a general framework for layer pseudospin superconductivity in moir\'{e} TMD materials, which deepens our understanding of the recently observed unconventional superconductivity in tMoTe$_2$, and also leads to several specific predictions.

We note that the main findings of our work are independent of the microscopic pairing mechanism, such as the Kohn-Luttinger mechanism that is a potential candidate~\cite{xu2025chiral}. Apart from the chiral pairing nature, the versatile diagram of finite momentum pairings exhibit the high tunability with external electric and magnetic fields. These enables the tMoTe$_2$ a potential platform for exploring finite-momentum topological superconductivity.

\section{Methods}

\textbf{Construction of layer-pseudospin model.}
\begin{figure}
		\centering
		\includegraphics[width=0.7\linewidth]{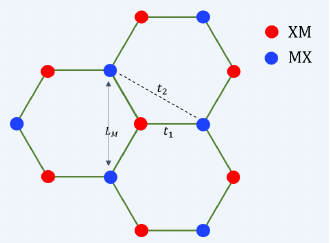}
		\caption{\textbf{Honeycomb lattice of layer pseudospin model.} The red/blue color denotes the XM/MX regions for the local stacking of twisted TMDs. }
		\label{fig:fig5}
\end{figure}

We construct the layer-pseudospin model by examining the generalized Kane-Mele model with sites centered on the honeycomb lattice formed by MX and XM stacking regions (see Fig.~\ref{fig:fig5}). The tight binding Hamiltonian takes the form~\cite{wu2019topological,devakul2021magic}
\begin{equation} \label{eq_effKaneMele}
\mathcal{H}_{\mathrm{tb}} = -t_1 \sum_{\langle i,j\rangle,\sigma} c^\dagger_{i\sigma} c_{j\sigma} +  t_2 \sum_{\langle \langle i,j\rangle\rangle,\sigma} e^{i\phi \sigma \nu_{ij}} c^\dagger_{i\sigma} c_{j\sigma},
\end{equation}
where $c^\dagger_{i\sigma}, c_{i\sigma}$ are fermionic creation/annihilation operators and $\sigma=\pm$ is the spin/valley degree of freedom. $\langle i,j\rangle$ and $\langle \langle i,j\rangle\rangle$ denotes the nearest and next-nearest-neighboring hopping, respectively. $\nu_{ij}=\pm 1$ depending on whether the next-nearest-neighboring hopping is anticlockwise ($+$) and  clockwise ($-$) with respect to the positive $z$ axis.

In the momentum space, we can directly find that $\gamma_0=t_2\sin\phi$ and $\gamma_2=t_2\cos\phi$ and $\gamma_1=t_1$. Therefore, the generalized Kane-Mele model can be rewritten as a layer-pseudospin model $H=\bm{d}\cdot\bm{\tau}$ due to the layer-polarized Wannier states (XM at top layer and MX at bottom layer)~\cite{devakul2021magic}. 

The mixing of interlayer and intralayer Cooper pairs can be analyzed by evaluating the overlap functions. For intravalley state, we have $d_1(\bm{k})=d_1(-\bm{k})$ and $d_2(\bm{k})=-d_2(-\bm{k})$ and $\sin (\theta(-\bm{k})/2)=\cos (\theta(\bm{k})/2)$. Using these relations, we can obtain
\begin{align}
&\langle u_{\bm{k},+}|\tau_x|u^*_{-\bm{k},+}\rangle=\cos^2\frac{\theta}{2}e^{i\alpha}+\sin^2\frac{\theta}{2}e^{-i\alpha} \\
&\langle u_{\bm{k},+}|\tau_0|u^*_{-\bm{k},+}\rangle=\sin\theta.
\end{align}
Therefore, the interlayer and intralayer Cooper pairs can be visualized on a layer Bloch sphere. We also get $\langle u_{\bm{k},+}|\tau_z|u^*_{-\bm{k},+}\rangle=0$ at zero displacement field, so we can only consider $\tau_0$ and $\tau_x$ channels. With the intralayer and interlayer attractive interaction $U$ and $V$, the full pairing susceptibility matrix is
	\begin{equation}
		\hat{\chi}(\bm{q},V_z,T)=\begin{pmatrix}
			\chi_{00}(\bm{q},V_z,T)&\chi_{01}(\bm{q},V_z,T)\\\chi_{10}(\bm{q},V_z,T)&\chi_{11}(\bm{q},V_z,T)
		\end{pmatrix}.\label{pairing_ma}
	\end{equation}
The superconducting transition temperature is determined when the maximal eigenvalue of $\hat{\chi}(\bm{q},V_z,T)-\mathrm{diag}(1/U,1/V)$ approaches zero.

\textbf{Continuum model of tMoTe2.} The moir\'{e} superlattice of twisted TMDs has a moir\'{e} lattice constant of $L_M=a_0/\sin\theta_t$, which folds the energy bands and  gives rise to the moir\'{e} Brillouin. $\theta_t$ is the relative angle between the two monolayer TMD. The continum model of twisted TMDs reads
	\begin{equation}\label{continuum}
		H_{\sigma}(\bm{r})=\begin{pmatrix}
			h_{b}(\bm{r})&\hat{T}(\bm{r})\\
			\hat{T}^{\dagger}(\bm{r})&h_t(\bm{r})
		\end{pmatrix},
	\end{equation}
where $\sigma=\pm$ is the spin/valley index for $\pm\bm{K}$ valley. Here the Hamiltonian of each individual layer is given by
	\begin{equation}
    h_{l}(\bm{r})=-\frac{1}{2m^*}(\hat{\bm{p}}-\sigma\bm{K}_{l})^2+\Omega_l(\bm{r})+l\frac{V_z}{2}.
	\end{equation}
Here $l=t(b)$ labels the top (bottom) layer and $m^*$ denotes the  effective mass of valence band. $\Omega_{l}(\bm{r})=\sum_{j=1,3,5}2V_m\cos(\bm{g}_j^{(1)}\cdot\bm{r}+l\psi)$ is  the intralayer moir\'e potential, and $\hat{T}(\bm{r})=w(1+e^{-i\sigma\bm{g}_2\cdot \bm{r}}+e^{-i\sigma\bm{g}_3\cdot \bm{r}})$ is the interlayer tunneling term. The reciprocal lattice vectors are $\bm{g}_i=\frac{4\pi}{\sqrt{3}L_M}(\cos\frac{(i-1)\pi}{3},\sin\frac{(i-1)\pi}{3})$. In our calculations we take $(m^*,\psi)=(0.62m_e,-105.9^\circ)$ and $(w,V_m)=(-18.8,16.5)$ meV adopted from Large-scale DFT simulations~\cite{xu2024maximally}. Here, $m_e$ is the bare electron mass.


\section{Data Availability}

The datasets generated during this study are available from the corresponding authors upon request.

\section{Code Availability }

The custom codes generated during this study are available from the corresponding authors upon request.

\section{Acknowledgements}
\begin{acknowledgments}

We thank Tingxin Li for inspiring discussions. K.T.L acknowledges the support from the Ministry of Science and Technology, China, and Hong Kong Research Grant Council through Grants No. 2020YFA0309600, No. RFS2021-6S03, No. C6025-19G, No. AoE/P-701/20, No. 16310520, No. 16307622, and No. 16309223. Y.M.X.  acknowledges financial support from the RIKEN Special Postdoctoral Researcher(SPDR) Program.

\end{acknowledgments}

\section{Author Contributions }
J.-X.H. and Y.-M.X. conceived the project idea. J.-X.H. performed the calculations and wrote the manuscript with input from Y.-M.X. and K.T.L. All authors contributed to the scientific discussions and manuscript revisions.

\section{Competing interests}
The authors declare no competing interests.

%
%

\end{document}